\documentclass[12pt]{iopart}

\usepackage{iopams}
\usepackage{bm}
\usepackage{graphicx}
\usepackage{amsfonts}
\usepackage{amssymb}
\usepackage{color}
\usepackage{verbatim}

\begin{document}
\title{Phase diagram of a frustrated asymmetric ferromagnetic spin ladder}

\author{Lihua Pan}
\address{School of Physics Science and Technology, Yangzhou University, Yangzhou 225002,China}
\author{Depeng Zhang}
\address{School of Physics Science and Technology, Yangzhou University, Yangzhou 225002,China}
\author{Hsiang-Hsuan Hung}
\address{Department of Physics, The University of Texas at Austin, Austin, Texas 78712, USA}
\author{Yong-Jun Liu}
\address{School of Physics Science and Technology, Yangzhou University, Yangzhou 225002,China}

\begin{abstract}
We perform a systematic investigation on the ground state of an asymmetric two-leg spin ladder (where exchange couplings of the legs are unequal) with ferromagnetic (FM) nearest-neighbor interaction and diagonal anti-ferromagnetic frustration using the Density Matrix Renormalization Group (DMRG) method. When the ladder is strongly asymmetric with moderate frustration, a magnetic canted state is observed between an FM state and a singlet dimerized state. The phase boundaries are dependent on the asymmetric strength. On the other hand, when the asymmetric strength is intermediate, a so-called spin-stripe state (spins align parallel on same legs, but antiparallel on rungs) is discovered, and the system experiences a first-order phase transition from the FM state to the spin-stripe state upon increasing frustration. We present numerical evidence to interpret the phase diagram in terms of frustration and the asymmetric strength.
\end{abstract}
\maketitle

\section{Introduction}
Frustrated quantum magnetism in low dimensions has received
considerable attention in recent decades. The exotic magnetic
behavior arises from  frustrated geometry, such as triangular
lattices and Kag\'ome lattices, which usually offer a large
degeneracy in the thermodynamic limt\cite{balents2010}. In quantum
limit, this is relevant to the quantum spin liquids. Recently,
ZnCu$_3$(OD)$_6$Cl$_2$, also called herbertsmithite, was
experimentally identified as a promising spin liquids and is modeled
as the spin-1/2 kag\'ome-lattice antiferromagnet\cite{han2012}.
Another indication for frustrated magnetism is in the spin-$1/2$
antiferromagnetic (AF) Heisenberg model on a two-dimensional square
lattice. Without frustration, the ground state is the N\'eel
ordering\cite{tangsanyee1989,manousakis1991} and the model provides
a description of the parents of
La$_2$CuO$_4$\cite{manousakis1991,dagotto1994}. With
next-nearest-neighbor (NNN) AF exchange, however, the other magnetic
collinear state will be triggered to compete with the N\'eel state.
As the NNN exchange is at the same order of the NN one, the ground
state is magnetically
disordered\cite{dagotto1989,schulz1996,richter2010}, and recently
has been identified to have a character of the spin liquids
state\cite{jiang2012}. The liquid state may be relevant to
high-temperature superconductors\cite{anderson1987,Leepa2006}.

The exotic state is however difficult to be determined without any
unambiguity. Theoretically the many-body effect prevents us against
directly studying it, and numerically the frustrated spin systems
suffer the minus sign problems, so the quantum Monte Carlo is unable
to accurately capture the behavior. The spin ladder instead provides
another route to study the physics since a ladder plays a role
crossover from one dimensions and two
dimensions\cite{white1994,dagotto1996,Uehara1996}. The frustrated
two-leg spin ladders have been intensively investigated during the
decade\cite{wangxiaoq2000,starykh2004,fath2001,Hung,hakobyan2007,ramakko2007,liugh2008,kimeugene2008,hikihara2010,dong2010,wenrui2011}.
Without frustrated NNN AF exchange, the ground state is singlet with
a finite spin gap\cite{white1994,barens1993}. In the large NNN AF
couplings limit, the ladder topologically behaves as a spin-1
chain\cite{fath2001}. In addition, it has been argued that there
exists an intermediate state, the columnar dimer state, in the
moderate frustration
regime\cite{starykh2004,liugh2008,hikihara2010}. This
indicates that even in a two-leg ladder frustrated magnetism can
provide rich phase diagrams.

Recently, the one-dimensional Heisenberg model with ferromagnetic
(FM) NN interaction with AF NNN exchange (represented as the zigzag
ladder) has also been frequently
mentioned\cite{Tonegawa89,Plekhanov,Bursill,Vekua,Meisner06,Meisner08,Hikihara,Sudan,KHida,jafari2007,Luht2006} since experimentally the
corresponding materials are discovered, including the edge-sharing
CuO$_{2}$ chains NaCu$_2$O$_2$\cite{Cu2}, quasi-onedimensional
helimagnet LiCuVO$_4$\cite{Cu3}, and powder curprate
Rb$_2$Cu$_2$Mo$_3$O$_{12}$\cite{Cu1} etc. In the frustrated FM
chain, an exotic phase with short-range incommensurate spin-spin
correlations is discovered by following the FM phase with moderate
frustration theoretically\cite{JSirker} and this exotic phase has been
characterized as Haldane dimer phase, which has a long-range order of
dimerization, and each dimer consists of spin-triplet pair\cite{Jap1,Jap2,Jap3}.
On the other hand, in
analogy to the zigzag ladder, the FM-leg and AF-rung ladder, also
indicates unconventional quantum magnetism
behavior\cite{Japaridze2007} and experimentally the corresponding
compound is also discovered\cite{Vekua,Yamaguchi2013}.

Next it is natural to consider a frustrated FM ladder, where both of
legs and rungs are FM, but diagonal exchange is AF. To the best of
our knowledge, such the frustrated FM ladder has not been drawn too
much attention. Therefore, we may ask whether or not there exists a
novel ground state phase diagram in the system. Even further, one is
wondering how much difference once the spin ladders are equipped
with inequivalent legs (asymmetric ladders). The leg asymmetry may
generate possible new phases and may induce different behaviors of
the excitation spectra in the limiting cases of weak and strong
asymmetries\cite{SChen,Essler,DNAristov}. The fascinating models
with geometric leg-asymmetry include the sawtooth strip
lattice\cite{sawtooth1,sawtooth2,sawtooth3,sawtooth4}, Comb-like
(necklace-like) model\cite{Essler} and the
diamond-chain\cite{diamond}. Therefore, theoretically it is
attractive to investigate the frustrated asymmetric FM ladders, in
particular, for highly asymmetric FM coupling in the legs and with
moderate AF diagonal frustration.

\begin{figure}
\begin{center}
\includegraphics[width=0.6\textwidth, bb=30 60 217 167]{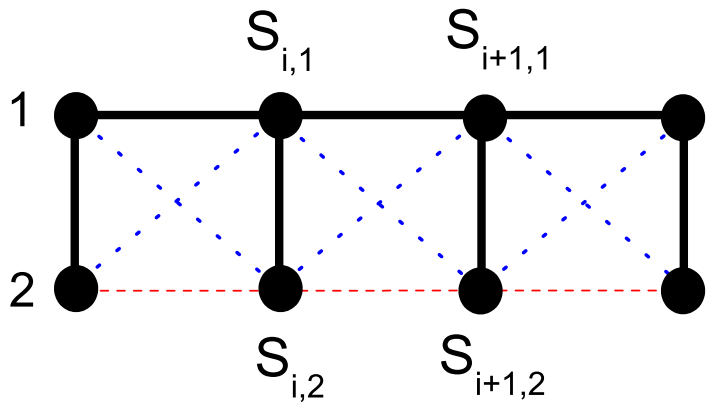}
\caption{(Color online) The frustrated asymmetric two-leg ladder.
The bold black lines denote the FM coupling, set as
unity, i.e. $J^1_{\parallel}=J_{\perp}=-1.0$. The red dashed line
denotes the asymmetric FM coupling $J^{2}_{\parallel}$. The blue
diagonal dotted lines are the AF exchange
$J_{\times}>0$.} \label{fig:fig1}
\end{center}
\end{figure}

The aim of this paper is to systematically study the quantum phase
transition (QPT) in the frustrated asymmetric FM spin ladder.
There are four ground state phases in the system: FM, partially polarized canted,  spin-stripe and singlet ($S=0$) dimerized states.
At weak asymmetry, increasing NNN AF coupling can induce a first-order transition from the FM
state to the spin-stripe state or to the dimerized state. At strong asymmetry, however, the processes between
the FM state to the canted state and  between the canted state to
the spin-stripe state are second-order. The singlet dimerized state
is characteristic of the $S=0$ state with dimerization in the strong frustration regions.
It is associated with the of two
neighboring sites spin correlations with alternating strengths.
This paper is arranged as follows. In Sec. \ref{sect:hamiltonian}, we
introduce the model Hamiltonian of the two-leg frustrated asymmetric
FM ladder. In Sec. \ref{sect:results}, we firstly present the ground state phase
diagram, then we introduce
the physical measurements used to distinguish the QPT using the
Lancz\'os diagonalization and the density matrix renomalization
groups (DMRG) method. Sec. \ref{sect:summary} is the summary.

\begin{figure}
\begin{center}
\includegraphics[width=0.7\textwidth, bb=0 0 495 568]{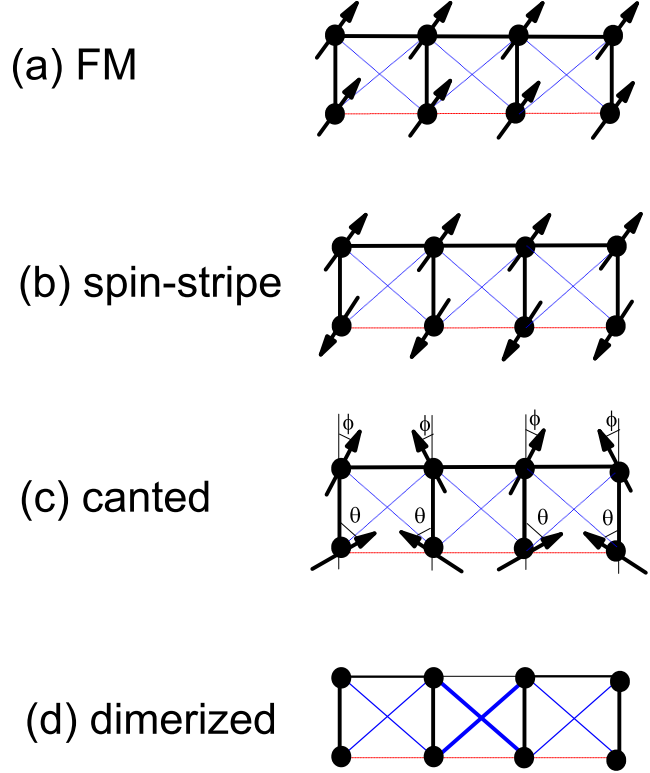}
\caption{(Color online) The classical pictures for (a) the FM state,
(b) the spin-stripe state, (c) the
non-collinear canted state. (d) shows the characteristic of the alternating strengths of the
spin correlations by different thickness of the bonds in the dimerized state.} \label{fig:fig2}
\end{center}
\end{figure}

\section{Model and method}
\label{sect:hamiltonian}

In this paper, we study the ground state phase diagram of the
 two-leg frustrated NN FM spin ladder with asymmetric legs, shown in
Fig. \ref{fig:fig1}. The model Hamiltonian reads as,
\begin{eqnarray}
H&=&J^1_{\parallel}\sum^{L-1}_{i=1}\vec{\mathbf{S}}_{i,1}\cdot
\vec{\mathbf{S}}_{i+1,1}+ J^2_{\parallel}\sum^{L-1}_{i=1}
\vec{\mathbf{S}}_{i,2}\cdot \vec{\mathbf{S}}_{i+1,2} \nonumber \\
&+& J_{\perp}\sum^L_{i=1} \vec{\mathbf{S}}_{i,1}\cdot
\vec{\mathbf{S}}_{i,2} \nonumber \\ &+&J_{\times}
\sum^{L-1}_{i=1}(\vec{\mathbf{S}}_{i,1}\cdot
\vec{\mathbf{S}}_{i+1,2}+\vec{\mathbf{S}}_{i,2}\cdot
\vec{\mathbf{S}}_{i+1,1}), \label{eq:hamiltonian}
\end{eqnarray}
where $\vec{\mathbf{S}}_{i,1}$ ($\vec{\mathbf{S}}_{i,2}$) denotes a
spin-$1/2$ operator of site-$i$ on the top (bottom) leg, and $L$ is
the length of the ladder. $J^1_{\parallel}$ and $J^2_{\parallel}$
are the intraleg couplings on the top and bottom leg,
respectively; $J_{\perp}$ denotes the interleg coupling on
the rungs. $J^{1,2}_{\parallel}$ and $J_{\perp}$ are all negative,
representing the NN FM exchange couplings.
In addition, we introduce the
diagonal AF exchange coupling $J_{\times}>0$ between the legs. The
existence of $J_{\times}$ leads spins to align anti-parallel and
competes with the $J^{1,2}_{\parallel}$, so it brings frustration.
For convenience, we hereafter set $J^{1}_{\parallel}=J_{\perp}=-1.0$,
indicated as the bold black lines in Fig. \ref{fig:fig1}.
The following calculations focus on the effects of asymmetric legs and the frustration.
The ratios $\alpha_{a} \equiv |J^2_{\parallel}/J^{1}_{\parallel}|$ and
$\alpha_{f} \equiv |J_{\times}/J^1_{\parallel}|$ are defined to
describe the asymmetric strength and the frustration strength,
respectively.

We perform the Lancz\'os
diagonalization \cite{linhq1990,Malvezzi} and the density matrix
renormalization group methods
(DMRG) \cite{Malvezzi,white1992,white1993,white1998,shibata2003,schollwock2005}
to study the ground state property. In contrast to the Lancz\'os
method, the DMRG is performed with a truncated Hilbert space where
the truncated basis is filtered by reduced density matrices
successively. Here the number of states is kept as $m=450$. We
denote $L$ as the ladder length so the number of sites is $N=2\times
L$. The size of the ladder in the DMRG calculation is up to $L=200$. We
exploited the conservation of the total magnetization $\sum_{i}S^z_{i}=0$ to
achieve higher precision of the calculations.
The truncation error is of the order of $10^{-7}$. For the DMRG
results, the open boundary conditions are used.


\section{Results}
\label{sect:results}

\begin{figure}
\begin{center}
\includegraphics[width=0.7\textwidth, bb=0 0 761 555]{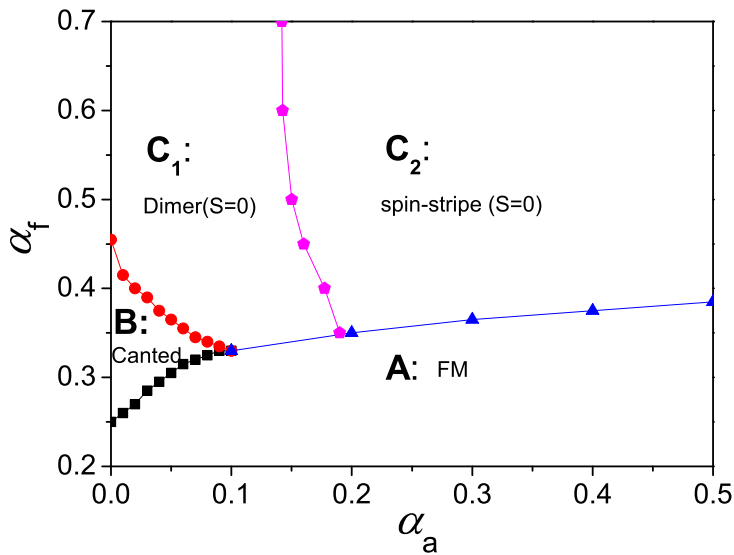}
\caption{(Color online) The ground phase diagram in the
$\alpha_{a}-\alpha_{f}$ plane, where $\alpha_a=|J^2_{\parallel}/J^1_{\parallel}|$ and $\alpha_f=|J_{\times}/J^1_{\parallel}|$ denote the asymmetry and frustration strengths, respectively. The FM state (phase {\bf A}) and canted state (phase {\bf B}) are magnetic. The dimerized (phase {\bf C$_1$}) and the spin-stripe
states (phase {\bf C$_2$}) are spin singlet with $S=0$. The phase
boundaries are numerically determined by finite-size analysis in the
thermodynamic limit $L=\infty$.} \label{fig:fig3}
\end{center}
\end{figure}

Let's start from the symmetric case where $J^{2}_{\parallel}=J^{1}_{\parallel}=-1.0$ and
consider the effects of frustration. It is well known that if there is no AF frustration, $J_{\times}=0$,
the ground state is a fully polarized FM state, where all spins
align in one direction, as shown in Fig. \ref{fig:fig2}(a). The totals spin is $S_{\textrm{max}}=2L s$, where $s=1/2$. On the
other hand, in the large $J_{\times}$ limit, the spins connected by
the diagonal AF bonds are anti-parallel aligned, and the ground state is singlet but has , i.e.
spin aligns parallel only along the leg direction, shown in Fig. \ref{fig:fig2}(b). The configuration of this spin-stripe state
has also been identified in a previous paper\cite{LinHQ}.
Therefore there exists a QPT with increasing
$J_{\times}$ (the frustration strength $\alpha_{f}$).

Then we consider a more generic case, the asymmetric ladder:
$J^1_{\parallel}\ne J^2_{\parallel}$.
Numerical simulations are performed to study the ground state phase diagram
in $\alpha_{a}-\alpha_{f}$ plane. When $\alpha_a$ is not too small ($\alpha_a\geq0.2$), the results are similar to
the symmetric case. The QPT from FM to spin-stripe is shown with increasing
$J_{\times}$, but the boundary location may be dependent on the value of $\alpha_a$.
The more fascinating results are shown at strong asymmetry (as $\alpha_a$ is
small). In addition to the FM and spin-stripe states, there exists a magnetic canted phase
and a dimerized state. In the canted state, the spin orientations are non-collinear [cf. Fig. \ref{fig:fig2}(c)]. Similar
to the {\it ferrimagnetism} of mixed spin
systems\cite{pati1997,wucongjun1999,ivanov2004}, magnetically it is
a partially polarized ordering state, and
the total spin is less than $S_{\textrm{max}}$. On the other hand, the dimerized state shows
a lone-ranged character of dimerization in spin-spin correlations along leg-1 and those along diagonal bonds  [cf. Fig. \ref{fig:fig2}(d)].

\begin{figure}
\begin{center}
  \includegraphics[width=0.7\textwidth, bb=0 0 366 357]{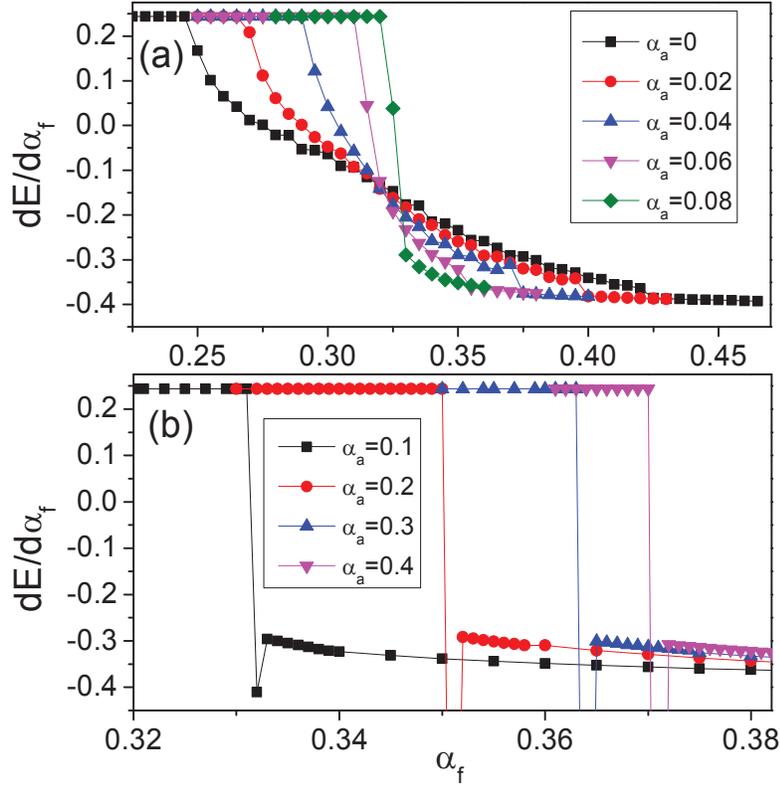}
\caption{(Color online) The first derivative of the ground state
energy $dE/d\alpha_{f}$ vs $\alpha_f$ for the $L=40$ ladder at different asymmetric
strengths $\alpha_{a}$.} \label{fig:fig4}
\end{center}
\end{figure}

\subsection{The ground state phase diagram}
A first-order QPT is characterized by a finite discontinuity in the
first derivative of the ground state energy. In the similar manner, a second-order QPT (or
continuous QPT) is characterized by a finite discontinuity
or divergence in the second derivative of the ground state energy;
here we assume that the first derivative is continuous\cite{book}.
Therefore, to detect the rank of QPT in the frustrated FM ladder,
one needs to calculate the ground state energy and its derivative.
In addition, we also measure the local magnetization (total spin per site)
and the spin correlation functions to identify
the ground state phase diagram.

The thermodynamic limit phase diagram  is presented in the
$\alpha_{a}-\alpha_{f}$ plane in Fig. \ref{fig:fig3}. The phase
are determined by using the DMRG method to perform
finite-size scaling. There exist four states, {\bf A}: FM state,
{\bf B}: the canted state, {\bf C$_{1}$}: the dimer $S=0$ state and
{\bf C$_{2}$}: the non-dimer $S=0$ spin-stripe state.
It is interesting to find that the canted state only exists
in the strong asymmetric case. At $\alpha_{a}=0$, the canted state
lies in the immediate frustration interval $0.25<\alpha_{f}<0.455$,
while the ground state is the FM state at $\alpha_{f}\leq0.25$ and the
dimer phase at $\alpha_{f}\geq0.455$. As $\alpha_{a}$ is lifted
off the zero onset, the canted state regime shrinks gradually. When
$\alpha_{a} \gtrsim 0.1$, the canted state vanishes and the transition
undertakes from the FM state directly  to the $S=0$ state upon
increasing frustration. In the following we will present numerical
evidence and analyze the corresponding pattern.

If the ground state is an FM state, the average energy per site can be
exactly determined as
\begin{eqnarray}E&=&\frac{1}{8}(1-\frac{1}{L})(J^{1}_{\parallel}+J^{2}_{\parallel}+2J_{\times})+\frac{1}{8}J_{\bot}, \nonumber\\
&=&\frac{1}{8}(1-\frac{1}{L})(2\alpha_f-(\alpha_a+1))J-\frac{1}{8}J,
\end{eqnarray}
where we consider open boundary conditions. Thus, in the FM state,
the first derivative of the ground state energy $dE/d\alpha_{f}$ is
a constant.

In Figs. \ref{fig:fig4}, we certainly observe the presence of the
plateau on $dE/\alpha_f$ in the small $\alpha_f$ regime. Thus we can
identify that phase {\bf A}  is the FM state. In both Figs.
\ref{fig:fig4} (a) and (b), the onsets for the plateau to collapse
moves towards to larger $\alpha_f$ with increasing $\alpha_a$.
Furthermore, Fig. \ref{fig:fig4} (a) shows that for $\alpha_a\lesssim 0.1$ the energy derivatives decrease smoothly, whereas (b) shows that for $\alpha_a >0.1$ they jump dramatically. This implies that, at small $\alpha_a$ (in the highly asymmetric case), the phase transition from the FM state to the phase with large $\alpha_f$ is second-order; otherwise it is first-order. We notice that the
occurrence of the constant $dE/d\alpha_f$ collapsed has weak
finite-size effect. Thus by small clusters, we can accurately
determine the phase boundaries from phase {\bf A} to phase {\bf B} or to
phase {\bf C$_{1,2}$}.

\begin{figure}
\begin{center}
  \includegraphics[width=0.7\textwidth, bb=0 0 765 541]{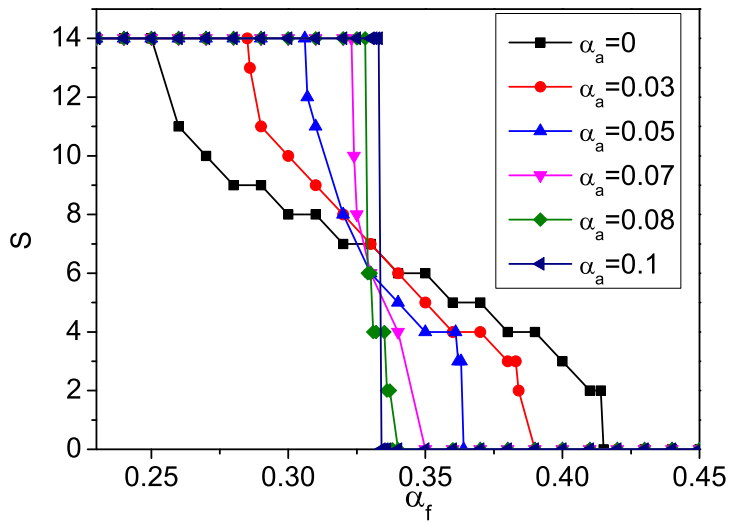}
\caption{(Color online) The total spin $S$ evolution as a function of
$\alpha_{f}$ at different asymmetric
strength $\alpha_{a}$ for the ladder system with length $L=14$
by Lancz\'os diagonalization.} \label{fig:fig5}
\end{center}
\end{figure}

\subsection{The total spin}
\begin{figure*}
\begin{center}
  \includegraphics[width=0.9\textwidth, bb=0 0 360 261]{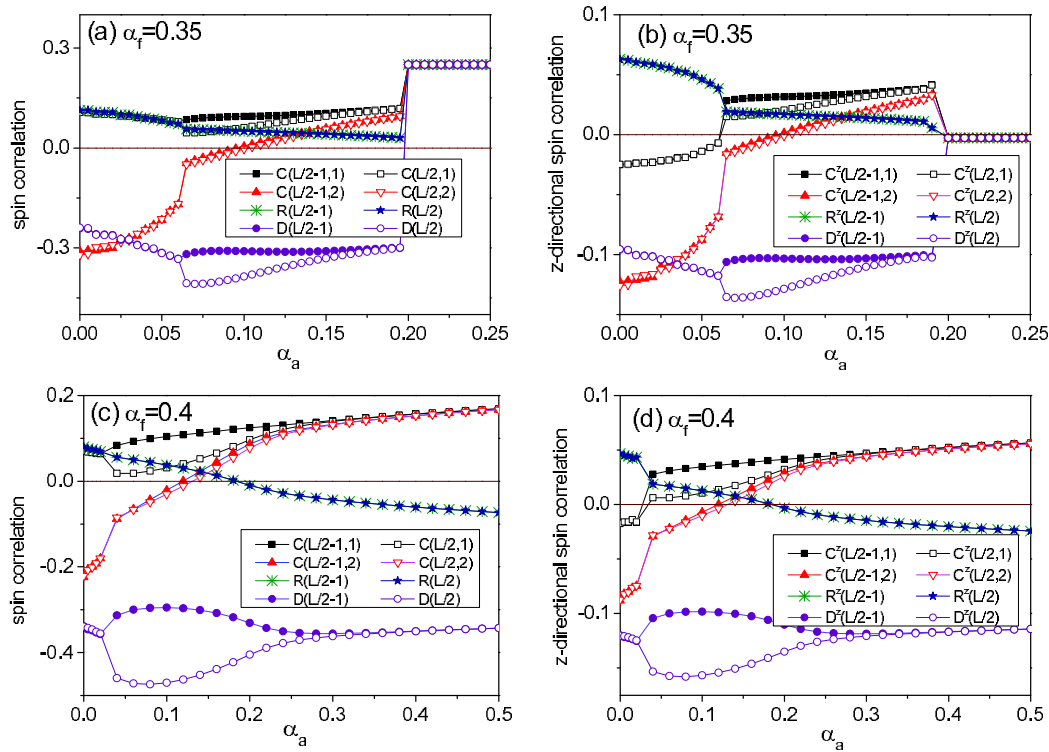}
\caption{(Color online) The neighbor spin correlation on the middle site $l=L/2$ (hollow) and it neighbor site $l=L/2-1$ (solid) with $L=48$ at (a-b) for $\alpha_f=0.35$ and
(c-d) for $\alpha_f=0.4$. (b) and (d) depict the spin correlations along $z$-direction.} \label{fig:fig6}
\end{center}
\end{figure*}
To further identify the magnetic nature of phase {\bf B} and {\bf C}, we
measure the total spin $S$. Physically, the total spin per site is the local magnetization. The total spin is defined as
\begin{eqnarray}
\langle \textbf{S}^2 \rangle=\sum_{ij}\sum_{ab}\vec{S}_{i,a}\cdot
\vec{S}_{j,b}=S(S+1).
\end{eqnarray}
Here we compute the total spin using Lancz\'os diagonalization, up
to $L=14$ with periodic boundary conditions. In Fig. \ref{fig:fig5},
we can see that at small $\alpha_f$, the value of $S$ remains on $S_{\textrm{max}}=Ns=L$, confirming that the ground state is fully polarized and is an FM state. Upon increasing $\alpha_f$ to a certain critical value, $S$ deviates from $S_{\textrm{max}}$, but in the small $\alpha_a$ regime,
the total spin still remains finite. Therefore the ground state is a magnetic state but is
partially polarized. This is similar to the ferrimagnetic
state \cite{pati1997,wucongjun1999,ivanov2004}, which has been
observed in mixed-spin systems (e.g. a spin-1/2-1 chain). In the classical analogy, we can identify that phase {\bf B} is
the canted state.

On the other side,  the magnetization  totally vanishes ($S=0$) at large $\alpha_f$. Thus both phase {\bf C$_{1,2}$} are non-magnetic.
However, the behaviors upon increasing $\alpha_f$ are markedly different for the strong and moderate asymmetric cases.
In the small $\alpha_a$ regime ($\alpha_a \lesssim 0.1$), increased $\alpha_f$ turns the FM state (phase {\bf A}) to the canted state (phase {\bf B}) and then to the singlet ($S=0$) state.
Note that this procedure is subject to the finite-size effect. Upon increasing the system size, the magnetization plateau will become smooth, and continuously decays to zero. As a consequence, the transition {\bf A}$\to${\bf B} and {\bf B}$\to${\bf C$_1$} can be classified as second-order. On the other hand, for $\alpha_a >0.1$, total spin directly drops to zero, and turns to the singlet states. This characterizes a first-order transition from {\bf A} $\to$ {\bf C$_{1,2}$}. The feature of these magnetic transitions is consistent with the previous analysis based on energy derivatives.

\subsection{The spin correlations}

\begin{figure*}
\begin{center}
  \includegraphics[width=0.9\textwidth, bb=0 0 360 261]{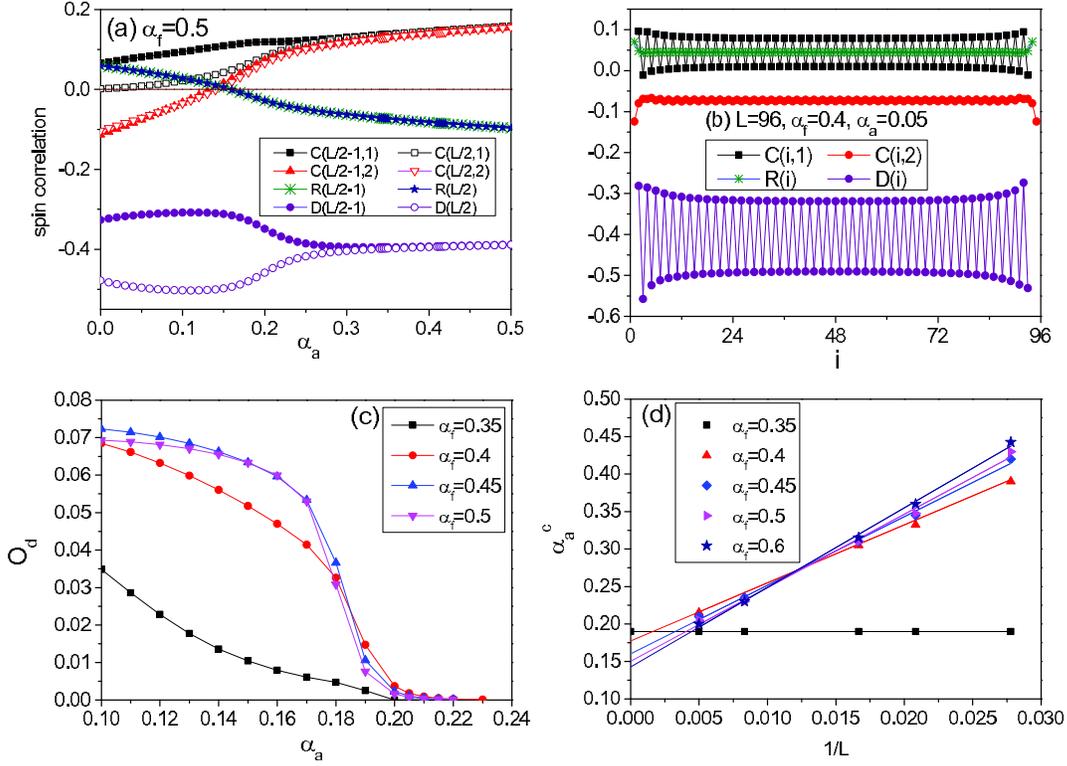}
\caption{(Color online) (a) The neighbor spin correlation on the middle site $l=L/2$ (hollow) and its neighbor site $l=L/2-1$ (solid) with $L=48$ for $\alpha_f=0.5$. (b) The neighbor spin correlations in the entire ladder system with $L=96$ for $\alpha_f=0.4$ and $\alpha_a=0.05$.
(c) the dimer order parameter as a function of $\alpha_a$ in the $L=200$ ladder but with different $\alpha_f$ values.
(d) gives the scaling calculations to locate critical transition point $\alpha_{a}^{c}$
between the dimer state and the non-dimer state in the thermodynamic limit. The line and the zero-point values
are get by linear fitting.
} \label{fig:fig7}
\end{center}
\end{figure*}

To further investigate the ground-state properties of the diagram, in this section,
we calculate the spin correlation functions.
The evolution of the spin correlation results as a function of $\alpha_a$ at different $\alpha_f$
can tell us that there must be several distinct phase regimes in the parameter space.
\begin{figure}
\begin{center}
\includegraphics[width=0.5\textwidth, bb=0 0 287 417]{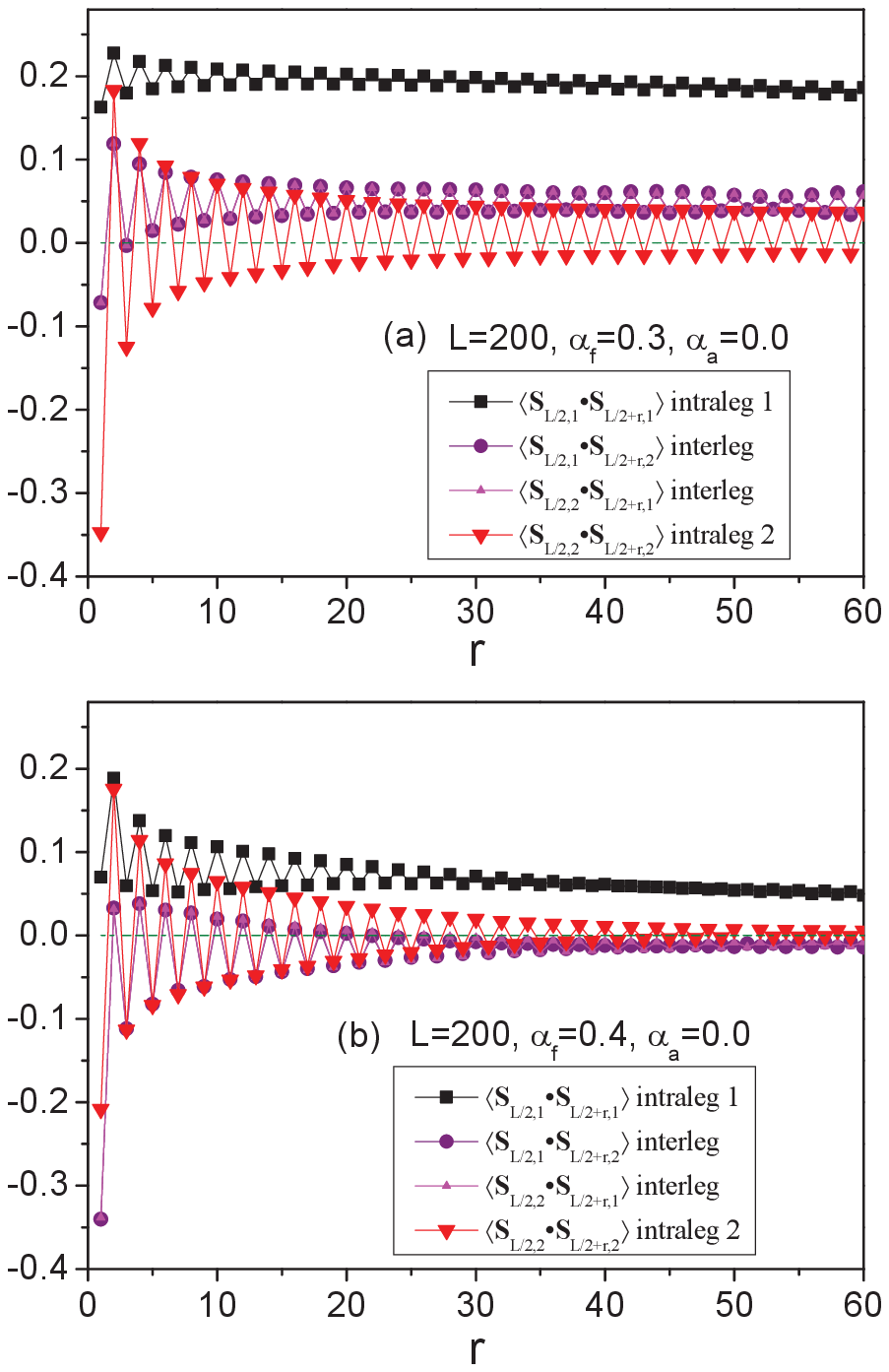}
\caption{(Color online) The long-range spin correlations of spins in the middle
rung with the other spin on the same leg and on the other leg, respectively,
as a function of the distance $r$. The parameter of the ladder considered is $L=200$ comb-like model($\alpha_a=0.0$) with different
frustration strength (a) $\alpha_f=0.3$, and (b) $\alpha_f=0.4$.
} \label{fig:fig8}
\end{center}
\end{figure}

Since our calculations are taken in the $\sum_{i}S^z_{i}=0$ subspaces,
the off-diagonal $xy$ and diagonal $z$ components of the spin correlations show different
behavior at different phase regimes. The magnetic symmetry breaking states, canted and FM states, show that
$\langle\vec{S}_{i}\cdot\vec{S}_{j}\rangle\neq3\langle S^z_{i} S^z_{j}\rangle$.
In the FM states, it can be easily deduced that all $\langle\vec{S}_{i}\cdot\vec{S}_{j}\rangle=0.25$, but the value of
 $\langle S^z_{i} S^z_{j}\rangle$ can be arbitrary.
However, all the zero-total-spin $S=0$ states preserve three-dimension rotational symmetry, so $\langle\vec{S}_{i}\cdot\vec{S}_{j}\rangle=3\langle S^z_{i} S^z_{j}\rangle$. For simplicity, the nearest neighboring spin correlations are denoted as $C(l,1)= \langle\vec{S}_{l,1}\cdot\vec{S}_{l+1,1}\rangle$, $C(l,2)= \langle\vec{S}_{l,2}\cdot\vec{S}_{l+1,2}\rangle$, $R(l)= \langle\vec{S}_{l,1}\cdot\vec{S}_{l,2}\rangle$ and $D(l)= \langle\vec{S}_{l,1}\cdot\vec{S}_{l+1,2}\rangle$. The corresponding correlations along $z$-direction are denoted as $C^z(l,1)= \langle S^z_{l,1} S^z_{l+1,1}\rangle$, $R^z(l) =\langle S^z_{l,1} S^z_{l,2}\rangle$ and $D^z(l) =\langle S^z_{l,1} S^z_{l+1,2}\rangle$, respectively.
In following correlation results, we choose the site index $l$ as the middlest site and the neighbor site, i.e. $l=L/2-1$ and $l=L/2$ on the $L=48$ ladder.

Figs. \ref{fig:fig6} (a) and (c) give the spin correlation results with moderate frustration, at $\alpha_f=0.35$ and $\alpha_f=0.4$, while
Figs. \ref{fig:fig6} (b) and (d) give their corresponding $z$-directional spin correlation results.

On the other hand, we can note the neighbor spin correlations in the regime with relatively weak asymmetry ($\alpha_a>0.4$) in Figs. \ref{fig:fig6}(c)-(d) and Fig. \ref{fig:fig7}(a), the ground states have zero total-spin $S=0$ and preserve the three-dimension rotational symmetry. It can be noted that when $\alpha_a$ is large ($\alpha_a \geq 0.4$), all $C(l,1)=C(l,2)>0$, indicating FM correlation along the leg direction.  On the other hand, $R(l)<0$ and $D(l)<0$ show AF correlation between the legs. Therefore, the system is a spin-stripe phase, where spins align parallel along the leg direction, as depicted in Fig. \ref{fig:fig2}(b). This state has zero spin excitation.
This observation is similar to the phase {\bf I} in the previous work.\cite{LinHQ}

Interestingly, in the $S=0$ state region, dimerization is detected at strong asymmetry.
In Figs. \ref{fig:fig6}(c-d) and Fig. \ref{fig:fig7}(a), where $0.02<\alpha_a<0.3$ and  $\alpha_a<0.35$, respectively, the marked amplitude differences between $C(L/2,1)$ (hollow) and  its neighbor counterpart $C(L/2-1,1)$ (solid) are observed. The dimerization patterns are also found in the diagonal correlations $D(L/2)$ and $D(L/2-1)$.
Furthermore, Fig. \ref{fig:fig7}(b) shows that the alternating strengths in spin correlations exist in the entire system.
However, the nearest neighboring spin correlations on the second leg $C(L/2,2)$ and $C(L/2-1,2)$ show uniform magnitudes, or much weaker alternating behavior.
These features are characteristic of the existence of  dimerization on the leg-1 in the case of strong asymmetry and frustration. The dimers in leg-1 are triplet dimers with FM correlations. This state has zero spin excitation even though the spin-spin correlations in this state has an exponential decay. We would like to mention that this is different from the gapped singlet dimer state\cite{liugh2008}.

To determine the boundary of regions where the dimer phase exists, we define the dimer order parameter as
$O_d=|\langle\vec{S}_{l,1}\cdot \vec{S}_{l+1,1}\rangle-\langle\vec{S}_{l+1,1}\cdot \vec{S}_{l+2,1}\rangle|$ to perform finite-size scaling analysis\cite{fath2001,Hung,Hung2011}.
We determine the phase boundary by observing the vanishing dimer order with the criterion: $O_d < 2\times10^{-4}$.
In Fig. \ref{fig:fig7}(c), it is obvious to see that as $\alpha_a$ increases in the $S=0$ regime, the dimer order parameter keeps decreasing, so there
exists a critical transition point $\alpha^c_{a}$ between the dimer $S=0$ state and the non-dimer $S=0$ state.
 Fig. \ref{fig:fig7}(d) shows the scaling behavior of critical transition points $\alpha^c_a$ at various $\alpha_f$ in the thermodynamic limit.

To further identify the spin arrangements of the canted states, we also calculate the long-range
the correlation function of spins under different distances $r$\cite{LinHQ}, i.e., $C(l,l+r)= \langle\vec{S}_{l,j}\cdot\vec{S}_{l+r,j'}\rangle$,
here, $j,j'=1$ ($2$) is corresponding to top (bottom) leg and $r$ is defined as the distance between the rungs where the two spins are located.
To avoid the boundary effect, we consider the inner spins of the finite chain by setting one spin in the middle rung $l=N/2$.
Figs. \ref{fig:fig8} give the long range correlation functions of spins in the middle rung with the other spin on the same leg and on the other leg, respectively,
as a function of the distance $r$.
From Fig. \ref{fig:fig8}(a), it can be seen that the spins on leg $1$ are FM correlated while the spins on leg $2$
are AF correlated, due to the strong leg asymmetry effect and the frustration effect. The long-ranged inter-leg spins are also FM correlated due to small
frustration $\alpha_f=0.3$. When the frustration strength is a little stronger, e.g., $\alpha_f=0.4$ in Fig. \ref{fig:fig8}(b), the long-ranged inter-leg
spins become AF correlated.
It can also be educed that the ferrimagnetic canted states for our model can not correspond to canted spiral or chiral orderings in
some zigzag ladders\cite{Jap1,Jap2,Jap3,zigzag,Japanchir} or
the leg spiral spin arrangements in a previous work \cite{LinHQ}.

\section{Summary}
\label{sect:summary} In summary, we have investigated the
ground-state phases of the frustrated asymmetric spin ladder with FM
nearest-neighbor interaction. In the strong asymmetric system, one
can clearly distinguish three phase regimes with frustration
increasing: from the pure FM state to canted state and then to $S=0$
dimer state. On contrary, when the asymmetric strength is not strong
enough , with frustration increasing there exists just one first
order phase transition from FM to spin stripe state.

\section{Acknowledgments}
This work is supported in part of NSFC Projects No. 11404281.
HHH acknowledges support from the Center for Scientific
Computing at the CNSI and MRL: an NSF MRSEC (DMR-1121053) and NSF
CNS-0960316.

\section{Author contribution statement}
All authors have contributed to the writing of the
manuscript and performing the calculations.

\end{document}